\documentclass[prl,twocolumn,superscriptaddress,showpacs,amssymb,amsmath,fixfloat]{revtex4}

\usepackage{graphicx}

\begin{document}
\title{The Persistence of High-Frequency Spin Fluctuations in Overdoped La$_{2-x}$Sr$_{x}$CuO$_{4}$ ($x$=0.22)}

\author{O.J. Lipscombe}
\affiliation{H.H. Wills Physics Laboratory, University of Bristol, Tyndall Ave., Bristol,
BS8 1TL, UK}
\author{S.M. Hayden}
\affiliation{H.H. Wills Physics Laboratory, University of Bristol, Tyndall Ave., Bristol,
BS8 1TL, UK}
\author{B. Vignolle}
\affiliation{H.H. Wills Physics Laboratory, University of Bristol, Tyndall Ave., Bristol,
BS8 1TL, UK}
\author{D.F. McMorrow}
\affiliation{London Centre for Nanotechnology and Department of Physics and Astronomy,
University College London, London, WC1E 6BT, UK}
\author{T.G. Perring}
\affiliation{ISIS Facility, Rutherford Appleton Laboratory, Chilton, Didcot, Oxfordshire
OX11 0QX, United Kingdom}

\begin{abstract}
We report a detailed inelastic neutron scattering study of the collective magnetic
excitations of overdoped superconducting La$_{1.78}$Sr$_{0.22}$CuO$_4$ for the energy
range 0--160 meV. Our measurements show that overdoping suppresses the strong response
present for optimally doped La$_{2-x}$Sr$_{x}$CuO$_4$ which is peaked near 50~meV. The
remaining response is peaked at incommensurate wavevectors for all energies investigated.
We observe a strong high-frequency magnetic response for $E \gtrsim 80$~meV suggesting
that significant antiferromagnetic exchange couplings persist well into the overdoped
part of the cuprate phase diagram.
\end{abstract}

\maketitle

The occurrence of high-temperature superconductivity is widely believed to be connected
to the cuprates' spin degrees of freedom \cite{Chubukov2003a}.  Thus, we might expect
strong spin fluctuations to co-exist with superconductivity over the whole
superconducting phase diagram.  The spin excitations have been well characterised for the
insulating antiferromagnetic (AF) \cite{Coldea2001} and lightly-doped
\cite{Keimer1992,Stock2006a,Mook2002a,Hayden2004} compositions. For optimally doped
compositions, structured excitations have been observed over a wide energy range
\cite{Hayden1996a, Vignolle2007a}. However, little is known about the high-energy spin
dynamics on the overdoped side of the cuprate phase diagram. One of the best materials to
investigate this region is single layer La$_{2-x}$Sr$_{x}$CuO$_{4}$ (LSCO).  This system
can be doped sufficiently to destroy the superconductive behavior \cite{Wakimoto2004},
allowing the spin excitations to be studied across the entire superconducting dome.
Studying the overdoped part of the phase diagram offers a different view on the emergence
of the superconducting state. In contrast to the underdoped regime, superconductivity
does not emerge from the `pseudogap' state \cite{Timusk1999}.  Rather, it emerges from
what appears to be a strongly correlated metallic state \cite{Hussey2003}.

\begin{figure}
\includegraphics[width=0.95\linewidth,clip]{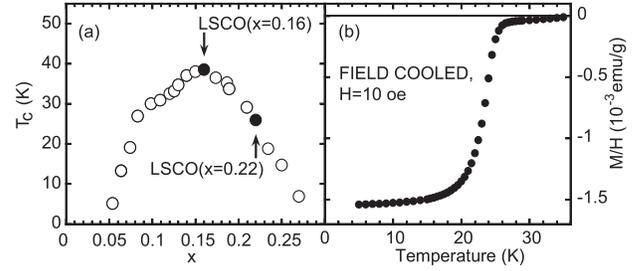}
\caption{(a) The doping dependence of $T_c$ in La$_{2-x}$Sr$_{x}$CuO$_{4}$. Open circles
\cite{Yamada1998}, closed circles: LSCO($x$=0.16) \cite{Vignolle2007a} and LSCO($x$=0.22)
(this work). (b) Magnetization of LSCO($x$=0.22) used in this work (H=10 oe $\parallel
c$, field cooled).} \label{Fig:phase_diagram}
\end{figure}

In this letter, we report an inelastic neutron scattering (INS) study of the magnetic
response $\chi^{\prime\prime}(\mathbf{q},\omega)$ of an overdoped superconducting sample
over a wide energy range (0--160~meV) and throughout the Brillouin zone. We have chosen
the composition La$_{1.78}$Sr$_{0.22}$CuO$_{4}$ ($T_c=26$~K), which shows a substantial
drop in $T_c$ with respect to optimal doping but is nevertheless superconducting (see
Fig.~\ref{Fig:phase_diagram}). We find that the spin excitations are dramatically
modified from those observed at optimal doping: The strong peak in the local
susceptibility $\chi^{\prime\prime}(\omega)$ present \cite{Vignolle2007a} in
La$_{1.84}$Sr$_{0.16}$CuO$_{4}$ near 50~meV is suppressed, and the remaining response is
incommensurate and strongest around $E \approx 10$~meV and $E \gtrsim 80$~meV. Thus,
strong spin excitations persist well into the overdoped region of the cuprate phase
diagram as required by magnetically mediated models of superconductivity.

La$_{1.78}$Sr$_{0.22}$CuO$_{4}$ has a tetragonal structure and we use tetragonal indexing
to label reciprocal space
$\mathbf{Q}=h\mathbf{a}^{\star}+k\mathbf{b}^{\star}+l\mathbf{c}^{\star}$. The magnetic
excitations in LSCO are 2D as the strongest magnetic couplings are within the CuO$_2$
planes.  Thus, usually we quote in-plane components of $\mathbf{Q}$.  In this notation,
the parent compound of the series, La$_{2}$CuO$_{4}$, exhibits AF order with an ordering
vector of $(1/2,1/2)$ and $\mathbf{a}^{\star}$ points along the Cu--O bonds. Seven single
crystals with a total mass of 75~g were co-aligned with a total mosaic of 0.8$^\circ$.
The crystals were grown by a traveling-solvent floating-zone technique \cite{Komiya2002a}
and annealed with one bar of oxygen for six weeks at 800$^\circ$C.  The Sr stoichiometry
was measured with SEM-EDX and ICP-AES to be $x=0.215 \pm 0.005$. Magnetization
measurements [see Fig.~\ref{Fig:phase_diagram}(b)] indicate that $T_c \mathrm{(onset)}=
26$~K.

\begin{figure*}
\includegraphics[width=0.85\linewidth,clip]{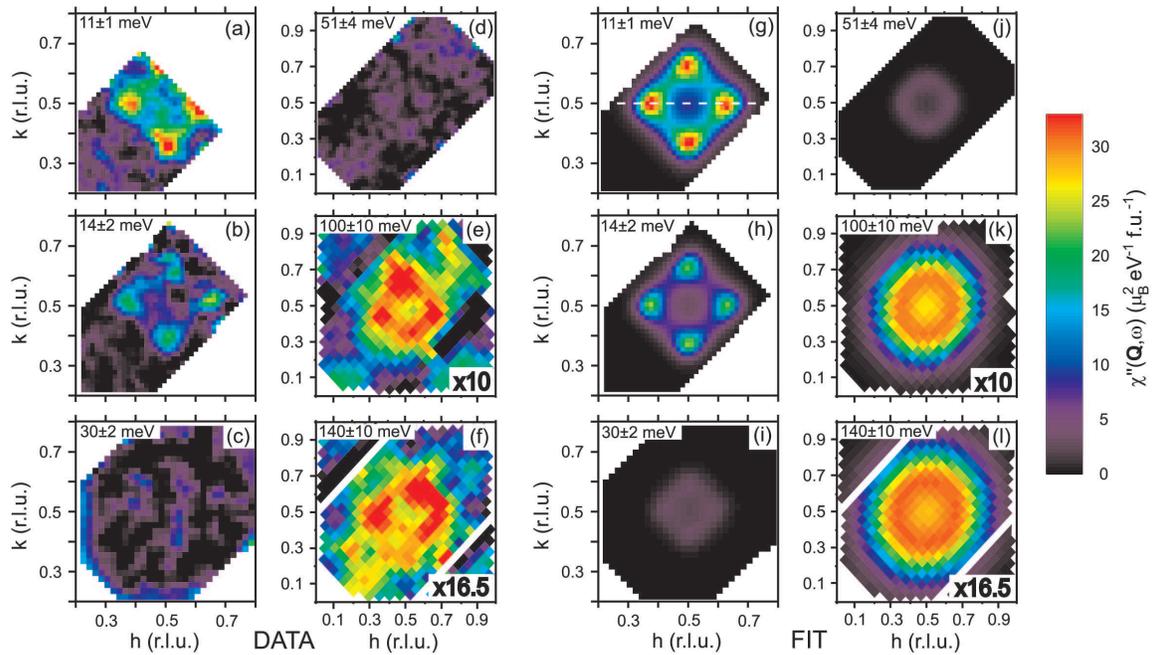}
\caption{(color online) (a)--(f) Constant-$E$ slices of $\chi^{\prime
\prime}(\textbf{q},\omega)$ for La$_{1.78}$Sr$_{0.22}$CuO$_{4}$ at $T=6$~K. Data are cut
off due to finite detector coverage. (g)--(l) Fits of model (Eq.~\ref{Eq:Sato_Maki})
convolved with instrument resolution. (a) and (b) show the low-$E$ four-peak structure
which is suppressed in (c) and (d). (e) and (f) show the reemergence of a ring-like
response at high energy. (a)--(d) have same intensity ranges, (e) and (f) are $\times10$
and $\times16.5$ respectively. $E_i$'s for (a)--(f) were 40, 40, 90, 120, 160, 240 meV.
\label{Fig:slices} }
\end{figure*}
INS probes the energy and wavevector dependence of $\chi^{\prime
\prime}(\textbf{q},\omega)$.  The magnetic cross section is given by
\begin{equation}
\label{Eq:cross_sect} \frac{d^2\sigma}{d\Omega \, dE} = \frac{2(\gamma
r_{\text{e}})^2}{\pi g^{2} \mu^{2}_{\rm B}} \frac{k_f}{k_i} \left| F({\bf Q})\right|^2
\frac{\chi^{\prime\prime}({\bf q},\hbar\omega)}{1-\exp(-\hbar\omega/kT)},
\end{equation}
where $(\gamma r_{\text{e}})^2$=0.2905 barn sr$^{-1}$, ${\bf k}_{i}$ and ${\bf k}_{f}$
are the incident and final neutron wavevectors and $|F({\bf Q})|^2$ is the anisotropic
magnetic form factor for a Cu$^{2+}$ $d_{x^{\scriptstyle 2}-y^{\scriptstyle 2}}$ orbital.
Data were placed on an absolute scale using a V standard.

The magnetic response of the cuprates is dominated by the unpaired $3d$ electrons of
Cu$^{2+}$ ions. The AF parent compound La$_{2}$CuO$_{4}$ displays spin waves
\cite{Coldea2001}, which disperse out from $(1/2,1/2)$ position.  For doped
superconducting compositions, the low-energy ($E \lesssim 20$~meV) excitations of
La$_{2-x}$Sr$_{x}$CuO$_{4}$ are peaked at $(1/2\pm\delta, 1/2)$ and $(1/2, 1/2\pm\delta)$
\cite{Shirane1989a,Cheong1991}. The low-energy incommensurability of the excitations
$\delta$ increases with $x$ and saturates with $\delta \approx 0.125$
\cite{Yamada1998,Wakimoto2004}. It has recently been shown that $\delta(E)$ of
La$_{2-x}$Sr$_{x}$CuO$_{4}$ and La$_{2-x}$Ba$_{x}$CuO$_{4}$ disperses
\cite{Tranquada2004,Christensen2004,Vignolle2007a} with energy and shows a minimum near
$E \approx 50$~meV. The high-energy ($E > 50$~meV) response in optimally doped
La$_{1.84}$Sr$_{0.16}$CuO$_{4}$ appears to be more isotopic, possibly with incommensurate
(IC) peaks rotated by $\pi/4$ in the $(h,k)$ plane \cite{Vignolle2007a}.  Interestingly,
a similar `hourglass' dispersion has been observed \cite{Arai1999,Bourges2000,Hayden2004}
in YBa$_{2}$Cu$_{3}$O$_{6+x}$.
\begin{figure}
\includegraphics[width=0.95\linewidth,clip]{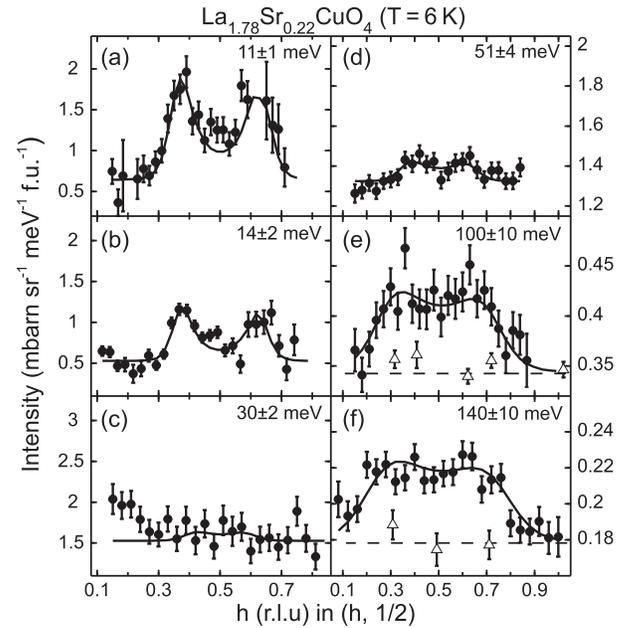}
\caption{(a)--(f) Constant-$E$ cuts through the raw data [see trajectory in Fig.
\ref{Fig:slices}(g)]. A background proportional to $\left| {\bf Q}\right|^2$ has been
subtracted. The energies are the same as the slices in Fig. \ref{Fig:slices}. Solid lines
are fits of Eq.~\ref{Eq:Sato_Maki} convolved with the instrument resolution. Background
points ($\triangle$) for (e)--(f) measured along ($h$,0.05). Scattering near $h$=0.2 in
(c) is due to phonons.} \label{Fig:6Kcuts}
\end{figure}

\begin{figure}
\includegraphics[width=0.9\linewidth,clip]{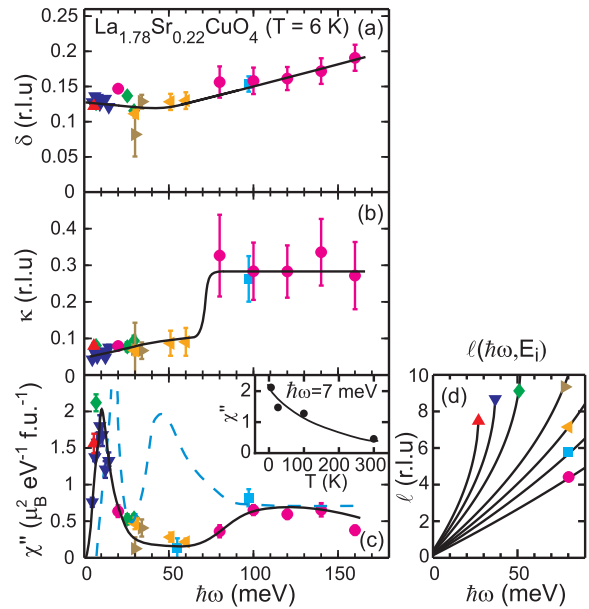}
\caption{(color online). (a)--(c) $E$-dependence of $\delta$, $\kappa$ and
$\chi^{\prime\prime}(\omega)$ in LSCO($x$=0.22). $\delta$, $\kappa$ define the
dispersion, and $\chi^{\prime\prime}(\omega)$ the strength of the spin excitations. The
inset shows how the response at 7 meV weakens on warming. Dashed line in (c) is
$\chi^{\prime\prime}(\omega)$ in LSCO($x$=0.16) for comparison \cite{Vignolle2007a}.
Lines are guides to the eye. (d) $E$-dependence of out-of-plane wavevector, $l$, at
$(1/2,1/2)$ for each $E_i$. Symbols indicate $E_{i}$: 30 ($\blacktriangle$), 40
($\blacktriangledown$), 55 ($\blacklozenge$), 90 ($\blacktriangleright$), 120
($\blacktriangleleft$), 160 ($\blacksquare$), 240 meV (${\bullet}$).} \label{Fig:params}
\end{figure}

Our INS experiments were performed on the MAPS instrument at the ISIS spallation source.
MAPS is a direct-geometry time-of-flight chopper spectrometer with position-sensitive
detectors. This allows a large region of reciprocal space to be sampled using a single
setting with a given incident energy $E_i$. In order to identify and minimize phonon
contamination of our results we collected data for seven different $E_i$'s. Each $E_i$
allows a given $(h,k,E)$ position to be probed for a different $l$ value [see Fig
\ref{Fig:params}(d)]. The magnetic scattering in LSCO depends only on the out-of-plane
wavevector $l$ through $\left| F({\bf Q})\right|^2$, whereas phonon scattering is
strongly dependent on $l$. We excluded data from our final results if it satisfied any of
the following criteria: (i) A clearly identified phonon branch could be seen to cross the
$(1/2,1/2)$ region. (ii) The pattern was not four-fold symmetric around $(1/2,1/2)$
(after subtracting a background quadratic in ${\bf Q}$). (iii) The fitted value of
$\chi^{\prime\prime}(\omega)$ was more than twice (including errors) that found at the
same $E$ for a different $E_i$. The excluding procedure was only required below the
phonon cut-off energy of 90 meV.

Fig. \ref{Fig:slices} shows $\chi^{\prime\prime}(\mathbf{q},\omega)$ of LSCO($x$=0.22) as
slices at various energies. Corresponding cuts through the raw data along the $(h,1/2)$
line are plotted in Fig.~\ref{Fig:6Kcuts}. The low-$E$ cut in Fig.~\ref{Fig:slices}(a) at
$E$=11 meV shows the well-known four-peak structure
\cite{Shirane1989a,Wakimoto2004,Cheong1991,Yamada1998,Tranquada2004,Christensen2004,Vignolle2007a}.
The structure is considerably weaker at $E$=14 meV [Fig.~\ref{Fig:slices}(b) and
Fig.~\ref{Fig:6Kcuts}(b)] and has almost disappeared for $E$=30 meV and $E$=50 meV
[Fig.~\ref{Fig:slices}(c)-(d) and Fig.~\ref{Fig:6Kcuts}(c)-(d)]. This behavior is in
contrast to optimally doped La$_{1.84}$Sr$_{0.16}$CuO$_{4}$
\cite{Christensen2004,Vignolle2007a}, where the IC response is strongest around $E
\approx 20$ meV and there is a strong response centered on $(1/2,1/2)$ for energies in
the range $E$=40--50 meV.  At higher energies, the magnetic response reemerges for
$E$=100 meV [Fig.\ref{Fig:slices}(e) and Fig.~\ref{Fig:6Kcuts}(e)] and $E$=140 meV
[Fig.\ref{Fig:slices}(f) and Fig.~\ref{Fig:6Kcuts}(f)]. The high-$E$ response is
significantly broader in $\mathbf{q}$ and $\chi^{\prime\prime}(\mathbf{q},\omega)$ is
weaker than the low-$E$ response. However, as we shall see below, when integrated in
$\mathbf{q}$ and $E$ it dominates the magnetic response.

We used a modified lorentzian function to make a quantitative analysis of the data:
\begin{equation}
\label{Eq:Sato_Maki} \chi^{\prime\prime}({\mathbf q},\omega)=\chi_\delta(\omega)
\frac{\kappa^4(\omega)} {[\kappa^2(\omega)+R(\mathbf{q})]^2}
\end{equation}
with
\begin{equation*}
R(\mathbf{q})=\frac{\left[(h-\frac{1}{2})^2+(k-\frac{1}{2})^2-\delta^2 \right]^2+\lambda
(h-\frac{1}{2})^2 (k-\frac{1}{2})^2} {4 \delta^2},
\end{equation*}
where the position of the four peaks is determined by $\delta$, $\kappa$ is an inverse
correlation length (peak width), and $\lambda$ controls the shape of the pattern
($\lambda$=4 yields four distinct peaks and $\lambda$=0 a pattern with circular symmetry
\cite{Lipscombe2007_note}). This phenomenological response function provides a good
description of the data at all energies. Fig.~\ref{Fig:slices}(g)-(l) show 2D fits of
Eq.~\ref{Eq:Sato_Maki} to the corresponding slices (a)-(f).  The parameters extracted
from fitting the resolution-convolved model to the 2D slices are shown in
Fig.~\ref{Fig:params}(a)-(c). We have expressed the strength of the magnetic response in
terms of the wavevector-averaged or local susceptibility
$\chi^{\prime\prime}(\omega)=\int \chi^{\prime\prime}(\mathbf{q},\omega) \; d^{3}q/\int
d^{3}q$ determined from the fitted $\chi^{\prime\prime}(\mathbf{q},\omega)$. The local
susceptibility indicates the overall strength of the magnetic excitations for a given
energy.

Fig.~\ref{Fig:params} summarizes the main findings of this work.  Firstly, the magnetic
response is made up of two components [see Fig.~\ref{Fig:params}(c)]: A low-frequency
(four-peaked) component for which $\chi^{\prime\prime}(\omega)$ is peaked around $E
\approx$ 10 meV and a much broader high-frequency component which is strongest around $E
\approx$ 120 meV. The spectral weight of the components are found to be $0.028(3)$ and
$0.053(4)$ $\mu^2_{B} \rm{f.u.}^{-1}$, the high energy component therefore dominating the
response. The emergence of the high-frequency component at $E \approx$ 80 meV corresponds
to a rapid broadening of $\chi^{\prime\prime}(\mathbf{q},\omega)$ in wavevector as shown
by the sudden increase in the $\kappa$ parameter [Fig. \ref{Fig:params}(b)] at this
energy. Comparing the present results with a recent study \cite{Vignolle2007a} of
optimally doped La$_{1.84}$Sr$_{0.16}$CuO$_{4}$ over the same energy range [dashed line
is Fig.~\ref{Fig:params}(c)], we note (i) The low-energy peak in
$\chi^{\prime\prime}(\omega)$ has moved from 18 meV down to about 10 meV with no change
in spectral weight \cite{Vignolle2007_note}. The low-energy IC component
[Fig.~\ref{Fig:params}(a)] does not disperse rapidly towards $(1/2,1/2)$ as seen in
optimally doped LSCO \cite{Christensen2004,Vignolle2007a}. Instead the pattern disappears
rapidly. (ii) The peak in $\chi^{\prime\prime}(\omega)$ around 40-60 meV present in
LSCO($x$=0.16), which corresponds to a strong response near $(1/2,1/2)$, is suppressed.
(iii) A high-energy ($E \gtrsim 80$~meV) component remains with approximately the same
amplitude as for LSCO($x$=0.16). A recent study \cite{Wakimoto2007} of more highly doped
LSCO($x$=0.25) reported significant spectral weight in the range 40-60 meV, although we
note that this work was performed with two $E_i$'s only, which may have limited the
ability to discriminate between scattering of a magnetic and phononic nature.

The origin of the magnetic excitations in the cuprates has been discussed in terms of
many models. As we move to the overdoped side of the phase diagram, it is widely believed
that Fermi-liquid-based models become more appropriate. Within such models, the
low-energy IC peaks arise from the creation of (correlated) quasiparticle pairs
\cite{Si1993,Littlewood1993,Norman2007a}.  Thus the shift of the low-energy peak in
$\chi^{\prime\prime}(\omega)$ to lower energy as we move from LSCO($x$=0.16) to overdoped
LSCO($x$=0.22) might be due to changes in the band structure.  It is more difficult to
explain the sudden collapse of the magnetic response in the intermediate 50--70 meV
energy range between LSCO($x$=0.16) and LSCO($x$=0.22) [see Fig.~\ref{Fig:params}(c)]
although photoemission (ARPES) suggests that the topology of the Fermi surface changes
between these two compositions as the quasiparticle states near $(1/2,0)$ move above the
Fermi energy \cite{Yoshida2006a}. Thus, there will undoubtedly be a concomitant change in
the nesting of the quasiparticle states. At higher energies (80--160 meV), we observe the
reemergence of a magnetic signal which despite the heavy doping is about 1/3 of the
intensity of the parent antiferromagnet La$_{2}$CuO$_{4}$ \cite{Hayden1996a} in the same
energy range.   If we associate the high-energy response with residual antiferromagnetic
interactions, we find from the dispersion of the high-energy excitations
[Fig.~\ref{Fig:params}(a)] $dE/d|\delta|$= $1800 \pm 500$~meV r.l.u.$^{-1}$ that $J=180
\pm 50$\ meV.  Thus a strong AF exchange coupling persists across the cuprate phase
diagram into the overdoped region.

It is interesting to compare our measurements with electronic spectroscopies. ARPES
experiments \cite{Kaminski2001a,Johnson2001a,Zhou2003a} have observed `kink' structures
in the quasiparticle dispersion of a number of systems including LSCO. These have been
interpreted as a signature of the coupling of quasiparticles to collective excitations
(bosons). In LSCO($x$=0.22) a kink is observed \cite{Zhou2003a} at about 70~meV which
corresponds approximately to the onset of the higher-energy excitations that we have
observed. Recent APRES experiments \cite{Kordyuk2006a,Graf2007a,Valla2007a, Chang2006a}
indicate further quasiparticle anomalies at higher energies which persist into the
overdoped region \cite{Kordyuk2006a}. These features could be related to high-energy spin
fluctuations which have been observed up to 250 meV for LSCO($x$=0.14) \cite{Hayden1996a}
and up to 160 meV in the present measurement of LSCO($x$=0.22). Infrared
optical-spectroscopy measurements \cite{Hwang2006a} also provide evidence of the coupling
of quasiparticles to bosonic excitations and in particular the existence of a
two-component excitation spectrum with a high-energy tail.

In conclusion, overdoping dramatically suppresses the magnetic response. The remaining
spin excitations are incommensurate and persist over a wide energy range. Defining the
energy scale of the spin excitations as the `center of mass' of the observed spectrum,
the parent compound is an antiferromagnet with spin-wave excitations and an energy scale
$E \approx$ $2J \approx 300$~meV. Doping results in stronger magnetic excitations at
intermediate energies, $E \approx 50$\ meV, for example, in optimally doped
La$_{2-x}$Sr$_{x}$CuO$_{4}$ and a lower energy scale. The present experiment shows that
the energy scale increases as we move into the overdoped regime.  Thus, the special
feature of the optimally doped region is the overall low characteristic energy of the
spin excitations and the drop in $T_c$ with overdoping is associated with the
disappearance of the commensurate response near 50 meV \cite{Vignolle2007a}.

%\bibliographystyle{aps5etal}
%\bibliography{highTc_bibtex}

\begin{thebibliography}{34}
\bibitem[{Lip(????)}]{Lipscombe2007_note}
 \bibinfo{note}{{The} data is best described with
  $\lambda=4(1-E/E_{\lambda})$ for $E < E_{\lambda}$ and $\lambda=0$ for $E
  \geq E_{\lambda}$, where $E_{\lambda}=60$~meV.}
\end{thebibliography}

\end{document}